\documentclass{elsarticle}
\pdfoutput=1

\journal{Nuclear Instruments and Methods}

\usepackage{amssymb}

\usepackage{url}
\usepackage{amsfonts}
\usepackage{amsmath}
\usepackage{epsfig}
\usepackage{lineno}
\usepackage[usenames]{color}
\usepackage{hyperref}
\usepackage{ulem}
\usepackage{graphicx}
\usepackage{caption}
\usepackage{subcaption}
\usepackage{array}
\usepackage{booktabs}
\usepackage{longtable}
\usepackage{chngcntr}

\newcommand{\EDrift}{$\text{E}_{\text{drift}}$}
\newcommand{\ETransfer}{$\text{E}_{\text{transfer}}$}
\newcommand{\EInd}{$\text{E}_{\text{ind}}$}
\newcommand{\EMMG}{$\text{E}_{\text{MMG}}$}
\newcommand{\VGEMOne}{$\Delta \text{V}^{\text{top}}_{\text{GEM}}$}
\newcommand{\VGEMTwo}{$\Delta \text{V}^{\text{mid}}_{\text{GEM}}$}
\newcommand{\VMMG}{$\text{V}_{\text{MMG}}$}
\newcommand{\CHFour}{$\text{CH}_{4}$}
\newcommand{\COTwo}{$\text{CO}_{2}$}
\newcommand{\CFFour}{$\text{CF}_{4}$}
\newcommand{\NTwo}{$\text{N}_{2}$}
\newcommand{\sigmaOverE}{$\sigma/\text{E}$}

\begin{document}

\begin{frontmatter}


\title{Combination of two Gas Electron Multipliers and a Micromegas as gain elements for a time projection chamber}

\author[yale]{S. Aiola}
\author[yale]{R.J. Ehlers}
\author[yale]{S. Gu}
\author[yale]{J.W. Harris}
\author[yale]{R. Majka}
\author[yale]{J.D. Mulligan}
\author[yale]{M. Oliver}
\author[texas]{J. Schambach}
\author[yale]{N. Smirnov}
\ead{nikolai.smirnov@yale.edu}
\address[yale]{Yale University, New Haven, CT, USA}
\address[texas]{University of Texas at Austin, Austin, TX, USA}

\begin{abstract}
We measured the properties of a novel combination of two Gas Electron Multipliers with a Micromegas for use as 
amplification devices in high-rate gaseous time projection chambers. The goal of this design is to minimize the 
buildup of space charge in the drift volume of such detectors in order to eliminate the standard gating grid and its
resultant dead time, while preserving good tracking and particle identification performance. We measured the 
positive ion back-flow and energy resolution at various element gains and electric fields, using a variety of 
gases, and additionally studied crosstalk effects and discharge rates. At a gain of 2000, this configuration
achieves an ion back-flow below 0.4\% and an energy resolution better than $\sigma/\text{E}=12\%$ for
$^{55}$Fe X-rays.
\end{abstract}


\end{frontmatter}


\section{Introduction}

A critical issue for time projection chamber (TPC) detectors is space charge distortion (SCD) due to 
the accumulation of positive ions in the TPC drift volume~\cite{VanBuren06}. This arises primarily from the ion 
back-flow (IBF) of positive ions from the gas amplification region, along with a contribution from primary ionization 
(from charged particles traversing the gas volume). Slow-moving positive ions distort the electric field 
uniformity and consequently distort the ionization electron drift trajectories, even for perfect external electric and 
magnetic field alignment and small transverse diffusion of the gas mixture.

The contribution of the primary ionization to the SCD can be minimized by two approaches. First, one can increase
the electric field in the TPC drift volume, as ion drift speed is approximately proportional to the electric field. Second, 
one can select a gas mixture to decrease the primary ionization itself, and to increase the ion 
mobility~\cite{Blum08}.

To minimize IBF, wire grid structures called gating grids (GGs) have traditionally been used~\cite{Amendolia85}. In 
the open state, GGs have a high transparency for ionization electrons to pass through to the gas amplification unit, 
typically a multi-wire proportional chamber. The GG can then be closed to collect ions from the gas 
amplification (gain) step. As a result, the IBF due to the gas amplification is very low. However, since the GG must remain closed 
until the positive ions from the avalanche at the anode wire have drifted to the grid, the TPC has an intrinsic dead 
time that limits the readout rate. Also, since the GG is a triggered element, there is a loss of track information near the 
readout planes during the time it takes to trigger and open the grid.

For current experiments employing large TPCs (e.g. STAR, ALICE) and those of the future, it is desirable to find a 
solution to minimize dead time by eliminating the GG or perhaps using a modified GG 
structure~\cite{Wieman14,Mulligan15}. The challenge is to minimize IBF from the gas amplification region to a level 
acceptable from the perspective 
of distortion corrections, such that track reconstruction and analysis have comparable performance to a GG 
solution~\cite{Sauli06,Colas04}. One possible solution is to use micro-pattern gas detectors (MPGDs), which 
have intrinsically low IBF. In particular, multi-layer MPGDs are promising candidates, as a stack of such 
elements allows multiple IBF-suppressing layers as well as flexibility in operational voltages and alignment, with only 
a small loss in electron transparency~\cite{Bondar03,ALICE14}. Simulations for the ALICE TPC~\cite{ALICE15} have 
shown that at the foreseen gain of 2000 (Ne+{\COTwo}+{\NTwo} (90--10--5)\footnote{This notation
reports the relative proportions of each gas in the mixture.}), with IBF as 
high as 2\% and energy resolution of 14\% ({\sigmaOverE}) or better (for $^{55}$Fe X-rays), TPC SCD can be corrected 
to an acceptable level in terms of TPC track finding, PID capability, and momentum resolution. In this paper, we report 
our investigation of the performance of a gain configuration for TPC gas amplification using two Gas Electron Multipliers 
(GEMs)~\cite{Sauli97} plus a Micromegas (MMG)~\cite{Giomataris96} in terms of IBF, energy resolution, and stability.

\section{Experimental Setup}

\begin{figure}[htb!]
\centering
\begin{subfigure}[t]{1.0\textwidth}
\includegraphics[width=\textwidth]{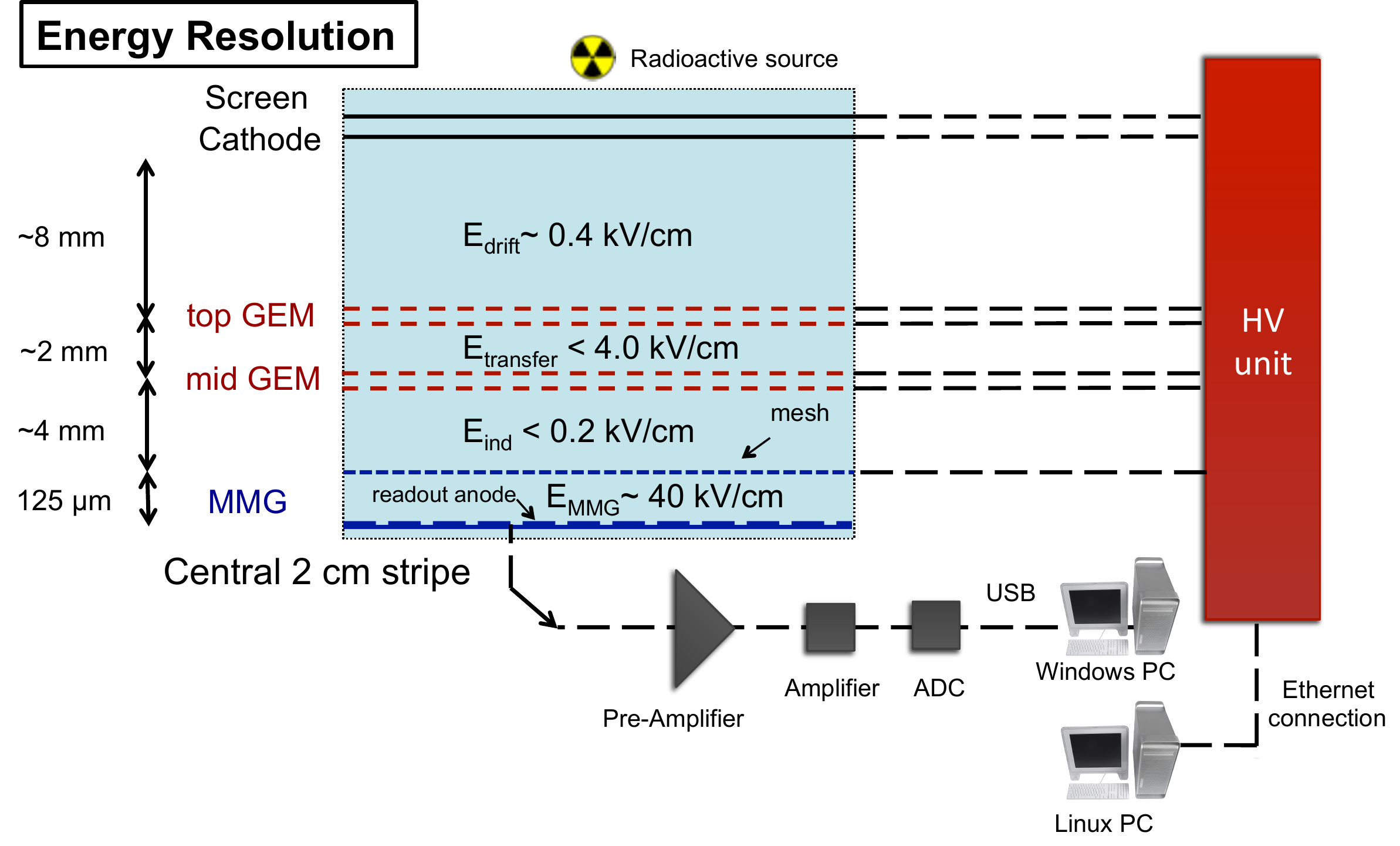}
\caption{Energy resolution measurement setup.}
\label{fig:setupERes}
\end{subfigure}

\begin{subfigure}[t]{1.0\textwidth}
\includegraphics[width=\textwidth]{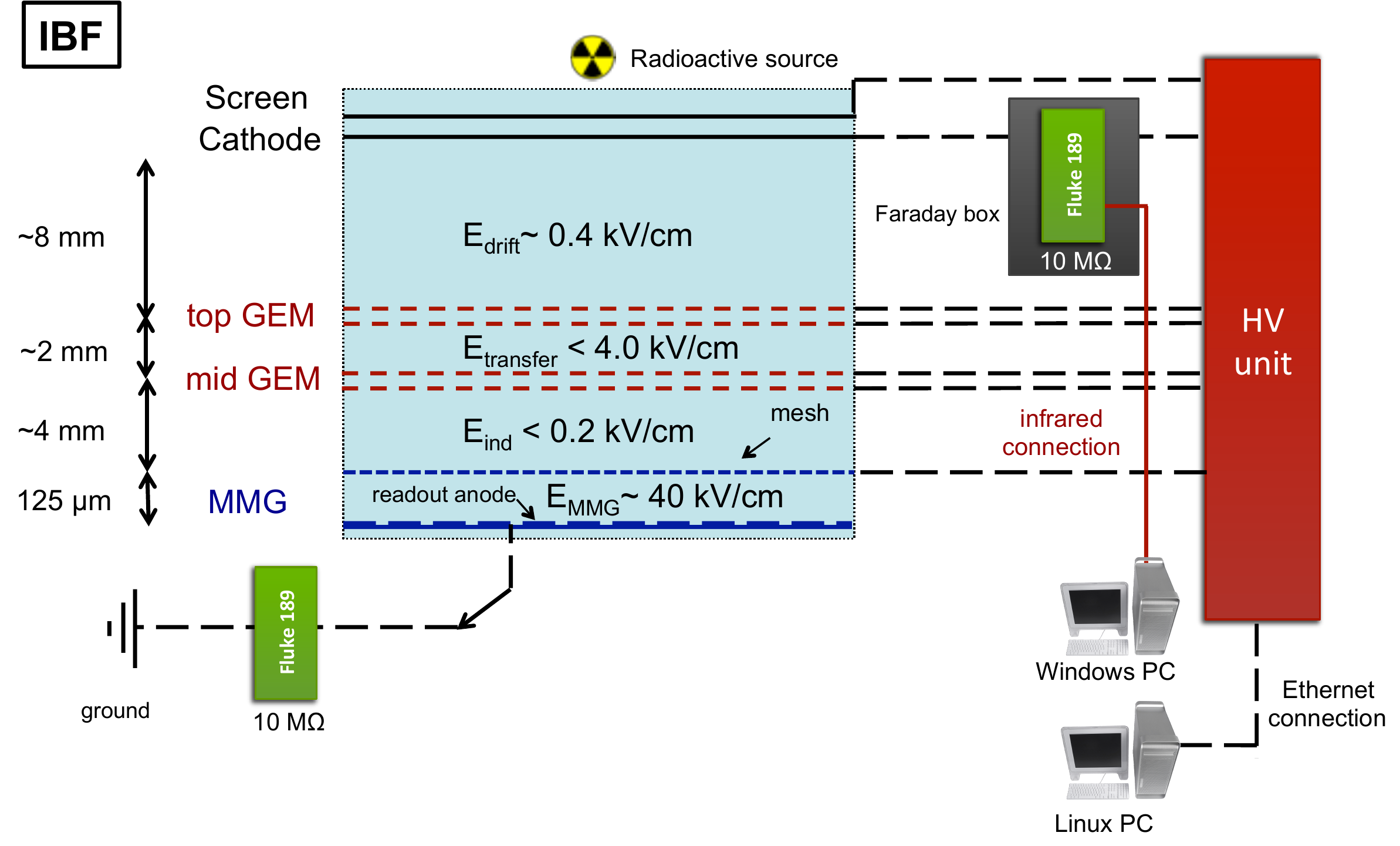}
\caption{IBF measurement setup.}
\label{fig:setupIBF}
\end{subfigure}
\caption{Experimental setup for a chamber with two stacked GEM foils and one MMG. The listed electric fields are the nominal values.}
\label{fig:experimentalSetup}    
\end{figure}

\begin{figure}[tbh]
\begin{center}
\includegraphics[width=\textwidth]{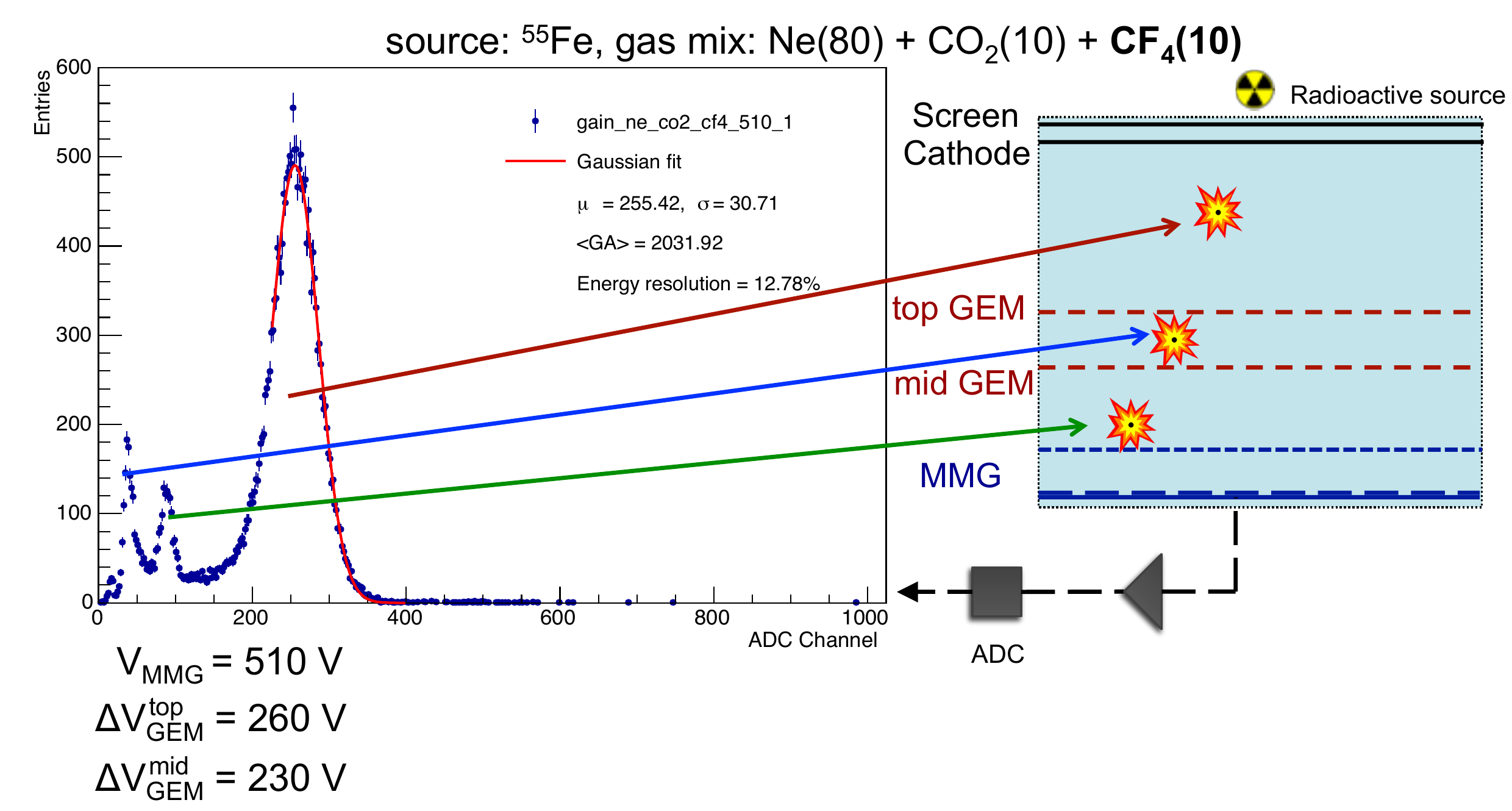}
\caption{An example of the $^{55}$Fe spectra showing the correspondence between the location of an X-ray absorption and each peak. }
\label{fig:exampleSpectra}
\end{center}
\end{figure}

Figure~\ref{fig:experimentalSetup} illustrates the 2-GEM + MMG setup used for these studies and defines the various 
elements and fields. The foremost operating principle is that the MMG provides most of the gain while the 
GEMs pre-amplify the signal for the MMG, so that it can be run at a relatively low voltage in order to reduce its discharge 
probability~\cite{Neyret12}. In addition, the GEMs help spread ionization electrons through diffusion and hole pattern misalignment 
so that a particularly dense cluster is less likely to cause a discharge in the MMG. 

The goal is then to tune the gains and fields in order to reduce IBF and increase energy 
resolution. The optimum effective gain of the GEMs is a compromise between better energy resolution, which would favor higher gain, 
and lower IBF from the GEMs, which would favor lower gain; the IBF contributed by a single GEM can be as much as 20\% of the ionization it produces. 
The top GEM is particularly sensitive to this trade-off, as it is the first gain element in the stack. Note that the effective 
GEM gain (total charge exiting the GEM divided by total charge drifting to the GEM) is a function of the voltage across the GEM, 
as well as the electric fields above and below the GEM~\cite{Sauli06}.
The IBF of the MMG scales with the ratio of the induction field to the MMG amplification field, 
{\EInd}/{\EMMG}~\cite{Colas04}, so the induction field is typically kept as low as possible. 
The primary purpose of the mid GEM is therefore to transfer electrons from the strong field in the transfer gap between 
the GEM foils to the lower field in the induction gap above the MMG. Accordingly, we operated the mid GEM with an effective 
gain less than 1. 
This feature can be seen clearly in the example spectrum shown in Fig.~\ref{fig:exampleSpectra}. In addition to tuning the voltages, the IBF can be 
further suppressed by arranging the GEM hole patterns to assure maximum mis-alignment. The top GEM foil was
rotated by 90$^{\circ}$ relative to the mid GEM to increase the hole mis-alignment.

This paper focuses on measurements of energy resolution and IBF under a variety of conditions in order to optimize 
performance of this detector design. We varied the voltages of each MPGD element ({\VGEMOne}, {\VGEMTwo}, and 
{\VMMG}), as well as electric fields between the elements ({\EDrift}, {\ETransfer}, and {\EInd}). We operated 
with a gas amplification 1500--2500, typical for a TPC readout in order to maintain a good signal 
to noise ratio with a reasonable electronic dynamic signal range for dE/dx measurements.
Additionally, we measured performance using a variety of gas mixtures, with 
argon and neon as the primary gases, and with {\COTwo}, {\NTwo}, {\CHFour}, and {\CFFour} as additional components. 
We conducted measurements with several small $10\times10\;\text{cm}^2$ chambers with a variety of readout plane 
geometries.\footnote{The first MMG was provided courtesy of L. Ropelewski, RD51, CERN.} There are two typical electronics 
configurations for these chambers: Fig.~\ref{fig:setupERes} shows the experimental 
setup for energy resolution measurements with the anode connected to a pre-amp/shaper amp/ADC chain.  
We measured the energy resolution as {\sigmaOverE} of $^{55}$Fe 
X-rays. To minimize electronic noise, we connected a small central section of the anode ($\approx20\ \text{cm}^2$) to the 
readout electronics, with the rest of the anode grounded. We used the $^{55}$Fe X-ray response in this configuration to set the chamber gain. 
Figure~\ref{fig:setupIBF} shows the setup for IBF measurements, where we connected the cathode 
to a high voltage source through a floating picoammeter while the anode was connected to a similar meter.

We maintained rather low operating voltages for all three gain elements and observed no discharges during these 
measurements. However, in a longer term experiment, discharges will occur. Thus, we performed all measurements 
with a discharge protection network at the preamplifier input, and took additional data (reported below) to estimate the discharge rate using 
both laboratory sources and a high intensity hadron beam.

\subsection{Measurement Procedure}

To characterize the performance with a given gas mixture, we first calibrated electronic gain by using a known capacitor and 
voltage step to inject a known charge into the preamplifier input. Then we took $^{55}$Fe spectra for several values of {\VMMG}. 
At each MMG setting, we tuned the GEM voltages to set the total chamber gain to $\approx2000$, and measured the energy 
resolution ($^{55}$Fe peak {\sigmaOverE}).  For each set of voltages, we then used an intense $^{90}$Sr source to measure 
the anode and cathode currents to calculate the IBF. For this measurement, we adjusted the source intensity to keep the 
anode current below 100 nA to avoid saturation from ion buildup in the chamber. For all measurements, the maximum 
water and oxygen content in the exit gas were 200 ppm and 30 
ppm, respectively, coming mainly from diffusion through the thin window in the chamber vessel.

Since the IBF currents are quite small, we took care to avoid noise and account for all current sources. As
seen in Fig.~\ref{fig:setupIBF}, we placed a screen electrode just outside the chamber cathode. The screen was operated
at the same voltage as the cathode and collects any ions produced outside the chamber. The picoammeter measuring the 
cathode current was placed in a shielded enclosure to avoid pickup noise and was read out by an infrared link to a computer. 
For each set of voltages, we measured the anode and cathode currents. In addition, we biased the chamber to measure the 
cathode current from the initial ionization in the drift gap and the anode current from ionization in the MMG induction gap. We checked 
the gain by approximating 

\begin{linenomath*}
    \begin{equation*}
	\left<\text{GA}\right>\approx\frac{\text{I}_\text{anode}- (\text{I}_\text{anode} \text{ from ionization in induction gap)}} 
	{\text{I}_\text{cathode} \text{ from ionization in drift gap}}.
    \end{equation*}
\end{linenomath*}

The IBF fraction for each voltage setting is then calculated as:

\begin{linenomath*}
    \begin{equation*}
        \text{IBF}=\frac{\text{I}_\text{cathode}- (\text{I}_\text{cathode} \text{ from ionization in drift gap})}
        {\text{I}_\text{anode} - (\text{I}_\text{anode} \text{ from ionization in induction gap)}}.
    \end{equation*}
\end{linenomath*}

The precision of all measurements is 10--15\% for IBF and 3--5\% for energy resolution. For IBF measurements, the dominant uncertainty
was due to pickup noise on the picoammeter. For energy resolution measurements, the dominant
uncertainty was systematic uncertainty in setting the fitting range in the Gaussian fitting procedure for the $^{55}$Fe peak. 

\section{Results}

\subsection{E-Field scans}

\begin{figure}[tb!]
\begin{center}
\includegraphics[width=\textwidth]{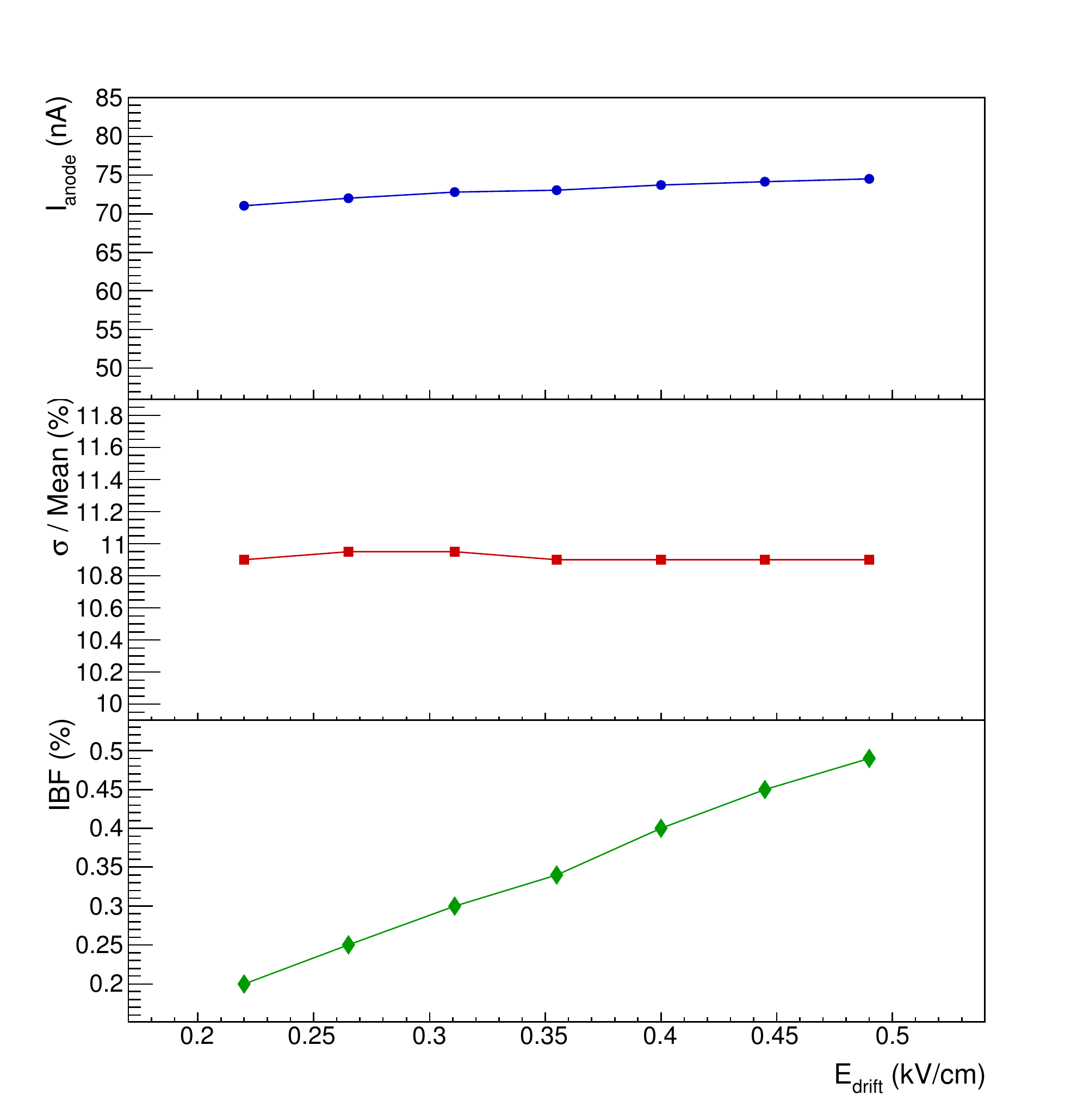}
\caption{Scan of {\EDrift} for Ne+{\COTwo} (90--10), {\VMMG} = 400 V, {\VGEMOne} = 242 V, {\VGEMTwo} = 185 V, {\ETransfer} = 2.0 kV/cm, {\EInd} = 0.075 kV/cm. Gain is $\approx2100$ (corresponding to $\text{I}_{\text{anode}} \approx74$ nA).}
\label{fig:NeCO2_EdriftScan}
\end{center}
\end{figure}

\begin{figure}[tbh!]
\begin{center}
\includegraphics[width=\textwidth]{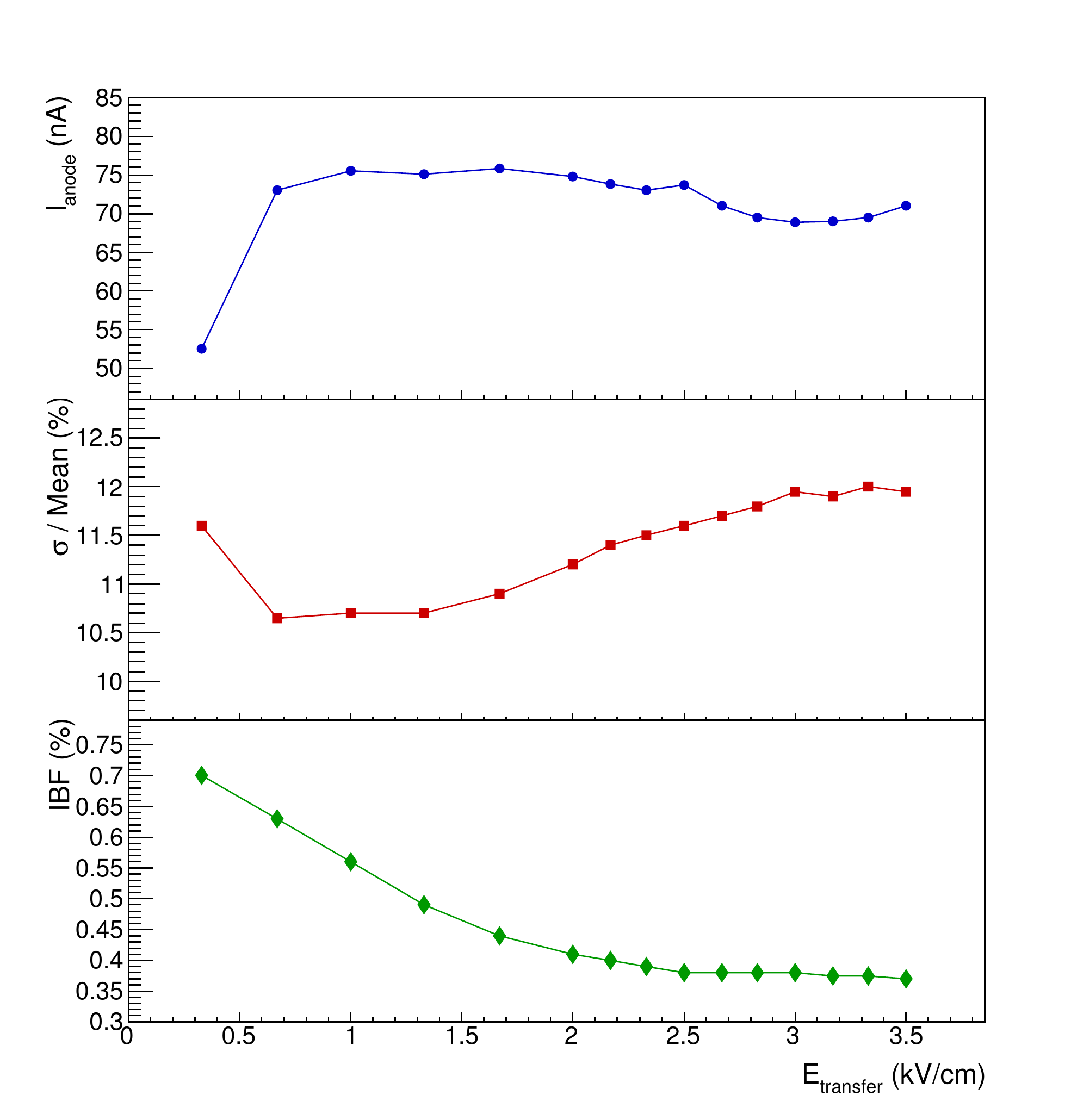}
\caption{Scan of {\ETransfer} for Ne+{\COTwo} (90--10), {\VMMG} = 400 V, {\VGEMOne} = 242 V, {\VGEMTwo} = 185 V, {\EDrift} = 0.4 kV/cm, {\EInd} = 0.075 kV/cm.}
\label{fig:NeCO2_EtranScan}
\end{center}
\end{figure}

\begin{figure}[tbh!]
\begin{center}
\includegraphics[width=1.\textwidth]{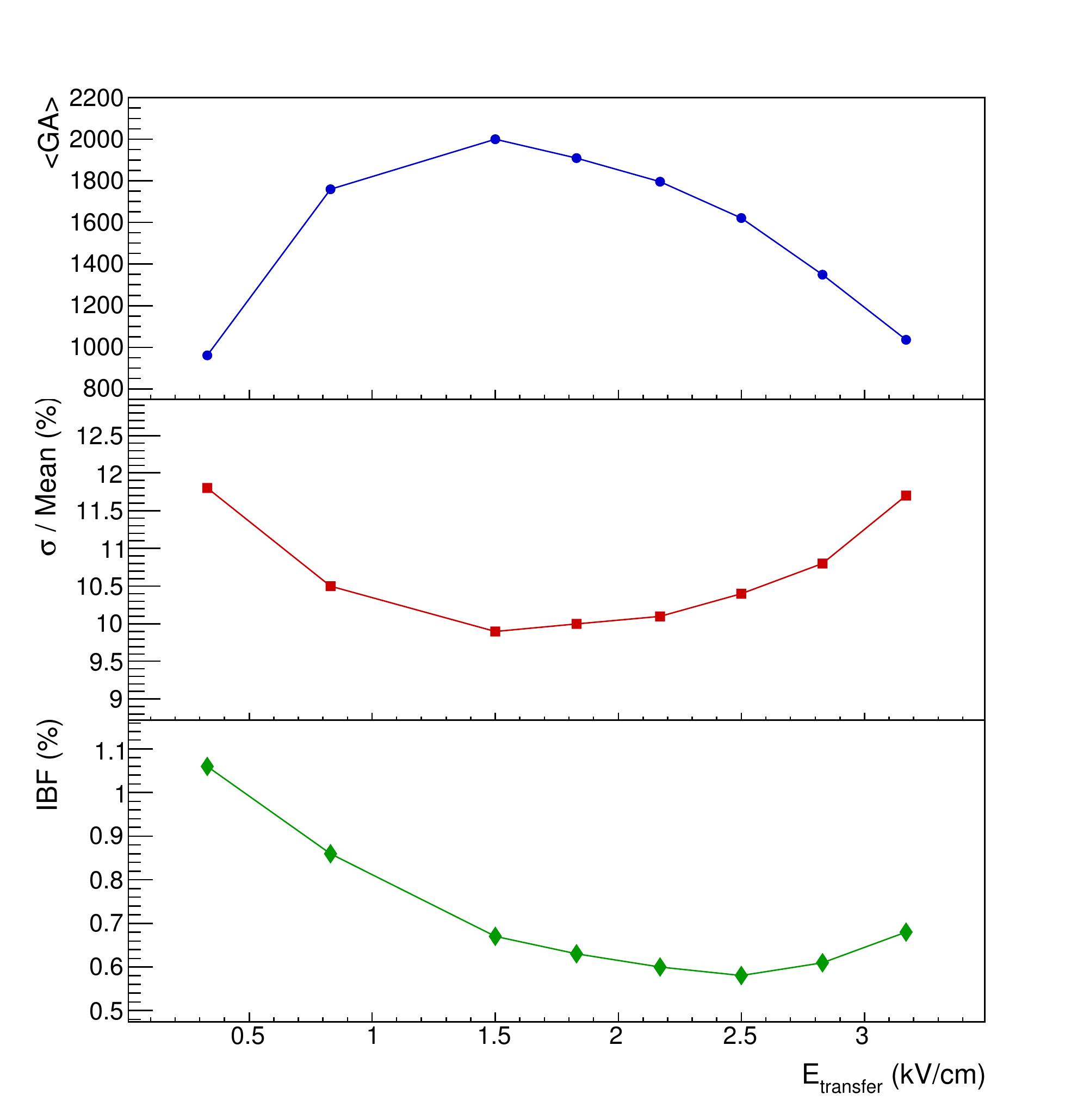}
\caption{Scan of {\ETransfer} for Ne+{\COTwo}+{\CFFour} (82--9--9), {\VMMG} = 430 V, {\VGEMOne} = 271 V, {\VGEMTwo} = 206 V, {\EDrift} = 0.4 kV/cm, {\EInd} = 0.075 kV/cm.}
\label{fig:NeCO2CF4_EtranScan}
\end{center}
\end{figure}

\begin{figure}[tbh!]
\begin{center}
\includegraphics[width=\textwidth]{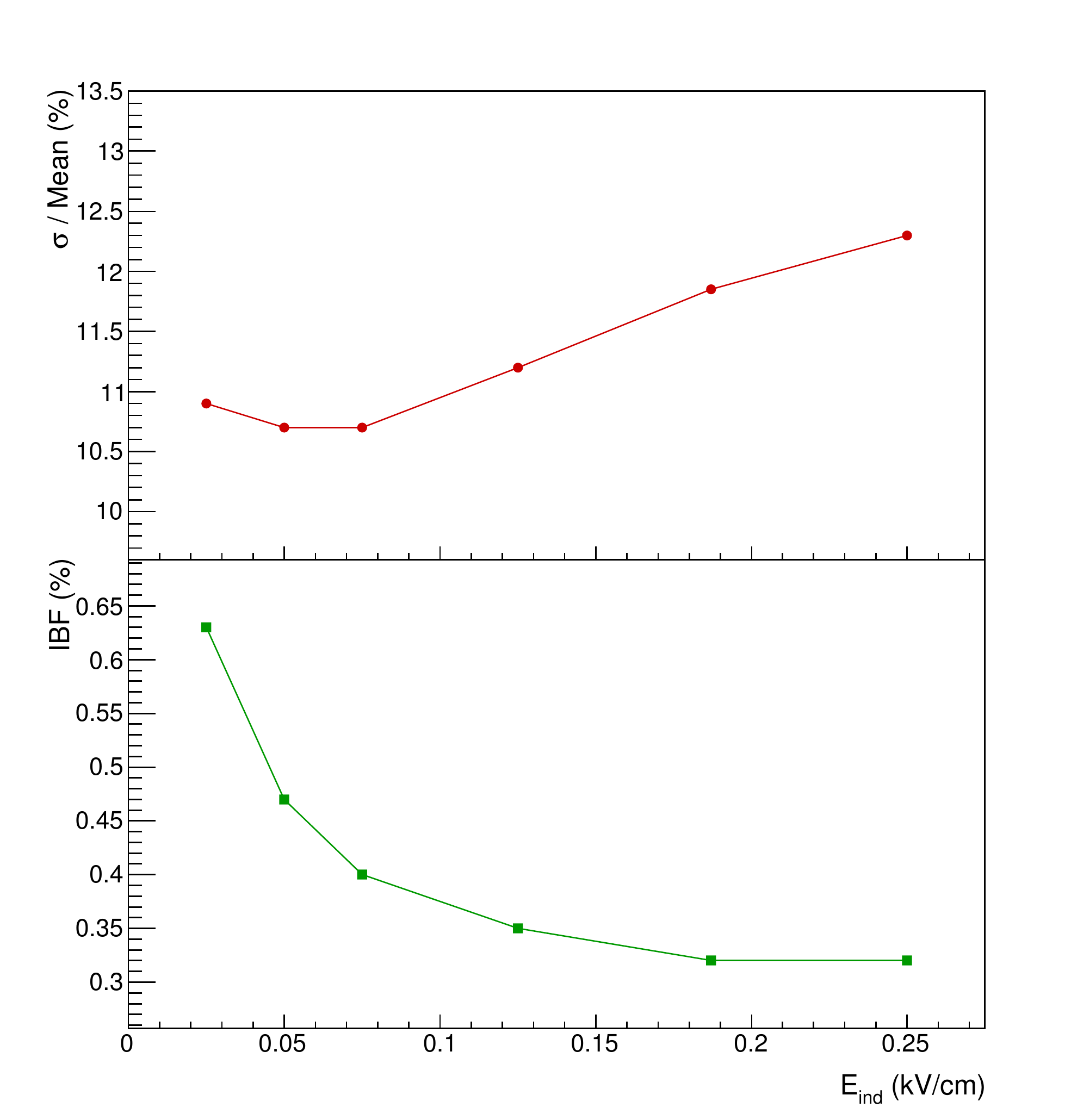}
\caption{Scan of {\EInd} for Ne+{\COTwo} (90--10), {\VMMG} = 400 V, {\EDrift} = 0.4 kV/cm, {\ETransfer} = 2.0 kV/cm. {\VGEMOne} and {\VGEMTwo} are tuned to keep the overall gain $\approx2000$.}
\label{fig:NeCO2_EindScan}
\end{center}
\end{figure}

Our first measurement characterized the detector performance as a function of {\EDrift}, {\ETransfer}, and {\EInd}; 
we performed field scans for each of the three fields in a Ne+{\COTwo} (90--10) gas mixture.

First, we varied {\EDrift} by changing the cathode voltage while keeping {\ETransfer} and {\EInd} constant. As {\EDrift} increases, ions
back-flowing from the top GEM are more likely to escape to the drift region. 
That is, the ion extraction efficiency increases. The IBF therefore depends almost linearly on {\EDrift}; doubling the field
approximately doubles the IBF (Fig.~\ref{fig:NeCO2_EdriftScan}, bottom panel). The anode current remains approximately constant as 
{\EDrift} increases; it is plotted to emphasize that the IBF trend
is not due to the small change in gain, but indeed due to the changing ion extraction efficiency from the top GEM 
(Fig.~\ref{fig:NeCO2_EdriftScan}, top panel). The energy resolution remains essentially constant through this scan
(Fig.~\ref{fig:NeCO2_EdriftScan}, middle panel), since the energy resolution depends weakly on the gain (the top GEM gain is
large enough to not statistically limit the resolution), and the gain changes weakly with {\EDrift} in the range studied. However, the operating
point for {\EDrift} is determined more by the drift requirements of the TPC than its effect on energy resolution and IBF; we therefore
operated at {\EDrift} = 0.4 kV/cm for all subsequent measurements.

Next, we scanned {\ETransfer} with {\EDrift} and {\EInd} fixed. As {\ETransfer} increases, the effective gain of the top GEM increases due to 
enhanced electron extraction efficiency~\cite{Sauli06}. This acts to improve the energy resolution, until it plateaus at {\ETransfer} $>1$ kV/cm. At the 
same time, the effective gain of the mid GEM decreases, which acts to degrade the energy resolution. The net effect is a balance between 
the behaviors of the two GEMs. The overall gain is fairly constant for {\ETransfer} $>1$ kV/cm (Fig.~\ref{fig:NeCO2_EtranScan}, top panel), 
and the energy resolution has a small degradation due to the decreased mid GEM gain (Fig.~\ref{fig:NeCO2_EtranScan}, middle panel). 
Moreover, the IBF improves as {\ETransfer} increases (Fig.~\ref{fig:NeCO2_EtranScan}, bottom panel)~\cite{Sauli06,Bondar03}. 
Nevertheless, within the limits $1.8<\;${\ETransfer}$<3.5$ kV/cm, there is only weak dependence of the 
energy resolution and the IBF on {\ETransfer}. Consequently, the operational {\ETransfer} should be in this vicinity. For 
{\ETransfer} $>3.7$ kV/cm in Ne+{\COTwo} (90--10), gas amplification begins to occur in the transfer region, setting an 
upper bound for {\ETransfer}. Accordingly, in the measurements below we operated with {\ETransfer}~$\approx2-3$ kV/cm.

Gas mixtures containing {\CFFour} have an additional constraint: for fields larger than 2.0 kV/cm, the gain decreases substantially 
due to resonant electron absorption by {\CFFour} (Fig.~\ref{fig:NeCO2CF4_EtranScan}). Note that in this scan the GEM voltages 
were not varied to keep the gain fixed. To avoid this absorption effect, we used {\ETransfer} = 1.5 kV/cm.

Finally, we scanned {\EInd} by fixing {\EDrift} and {\ETransfer} (as well as {\VMMG}), and tuning the GEM 
voltages to preserve the gain $\approx2000$.  Similar to the case of the {\ETransfer} scan, increasing {\EInd} increases 
the electron extraction from the mid GEM, which increases the gain. The GEM voltages are decreased accordingly 
to keep the gain constant. In particular, as {\EInd} increases, 
decreasing the top GEM gain results in a degradation of the energy resolution, and a decrease of the IBF 
(Fig.~\ref{fig:NeCO2_EindScan}). We chose to work with {\EInd} = 0.075 kV/cm in all subsequent measurements.

\subsection{Energy Resolution vs. IBF}

\begin{figure}[htb!]
\begin{center}
\includegraphics[width=1.\textwidth]{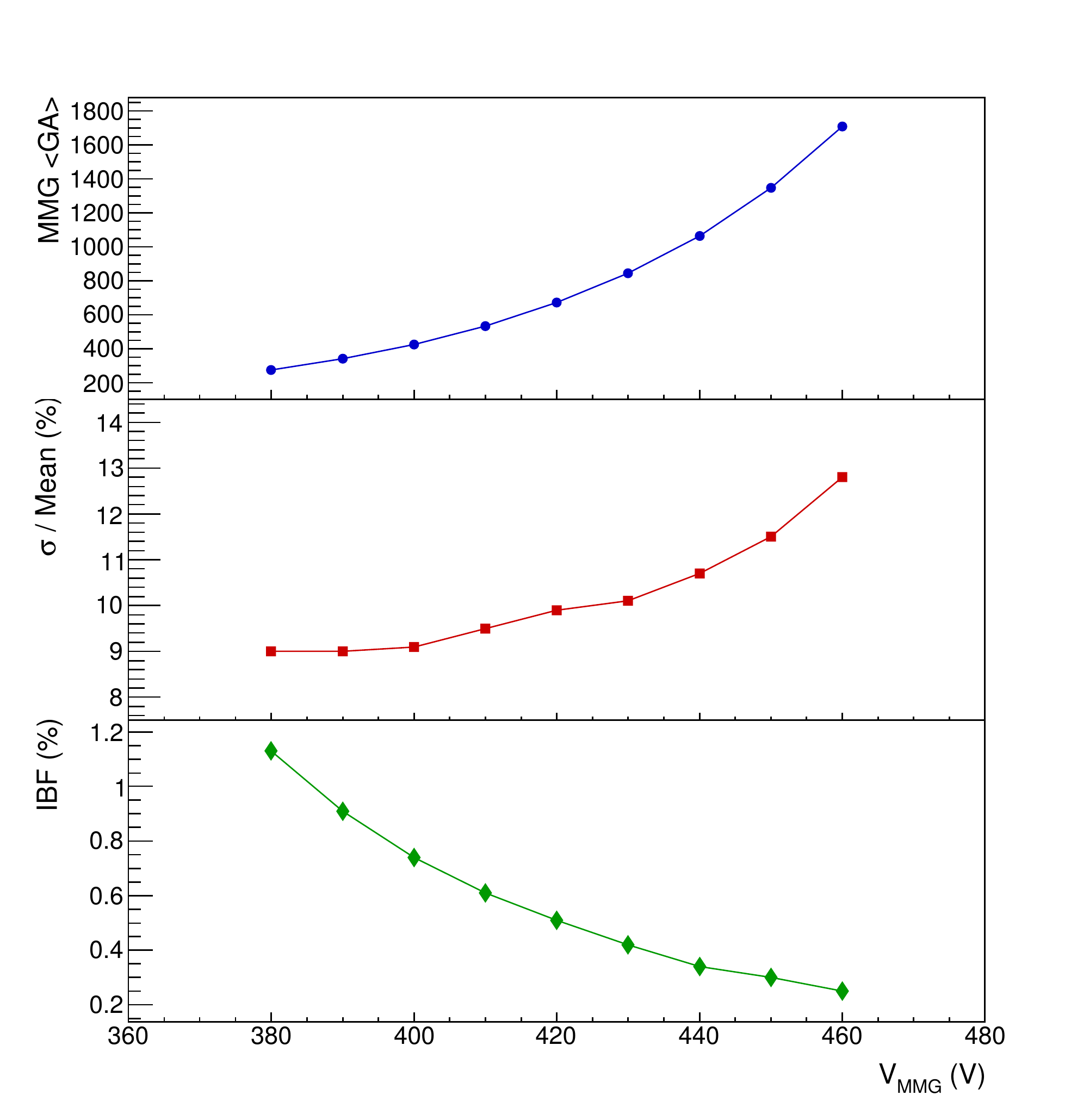}
\caption{MMG gain, energy resolution, and IBF as a function of the {\VMMG} for Ne+{\COTwo}+{\CHFour} (82--9--9), {\EDrift} = 0.4 kV/cm, {\ETransfer} = 3.0 kV/cm and {\EInd} = 0.075 kV/cm. {\VGEMOne} and {\VGEMTwo} are tuned to keep the overall gain $\approx2100$.}
\label{fig:NeCO2-one-scan}
\end{center}
\end{figure}

Next, we studied how to optimally distribute gain through the three elements in terms of maintaining good energy resolution and IBF. 
We changed {\VMMG} in steps of 10 V, starting at a voltage corresponding to a MMG gain of 200, 
and then tuned the GEM voltages to preserve the overall gain of about 2000. 
Throughout the measurements, we fixed {\EDrift}, {\ETransfer}, and {\EInd} at 0.4, 3.0, and 0.075 kV/cm, respectively. 
Fig.~\ref{fig:NeCO2-one-scan} illustrates the results of such a set of measurements.
As discussed previously, when the MMG gain is smaller (with correspondingly higher gain in the GEMs), the energy
resolution improves (Fig.~\ref{fig:NeCO2-one-scan}, middle panel) at the expense of a higher IBF (Fig.~\ref{fig:NeCO2-one-scan}, bottom panel). At the other extreme,
when the gain is almost entirely provided by the MMG, IBF improves at the expense of worse energy resolution.
Thus, the IBF and energy resolution anti-correlate with each other when the gas amplification share of each gain
element is varied.

\begin{figure}[htb]
\begin{center}
\includegraphics[width=1.\textwidth]{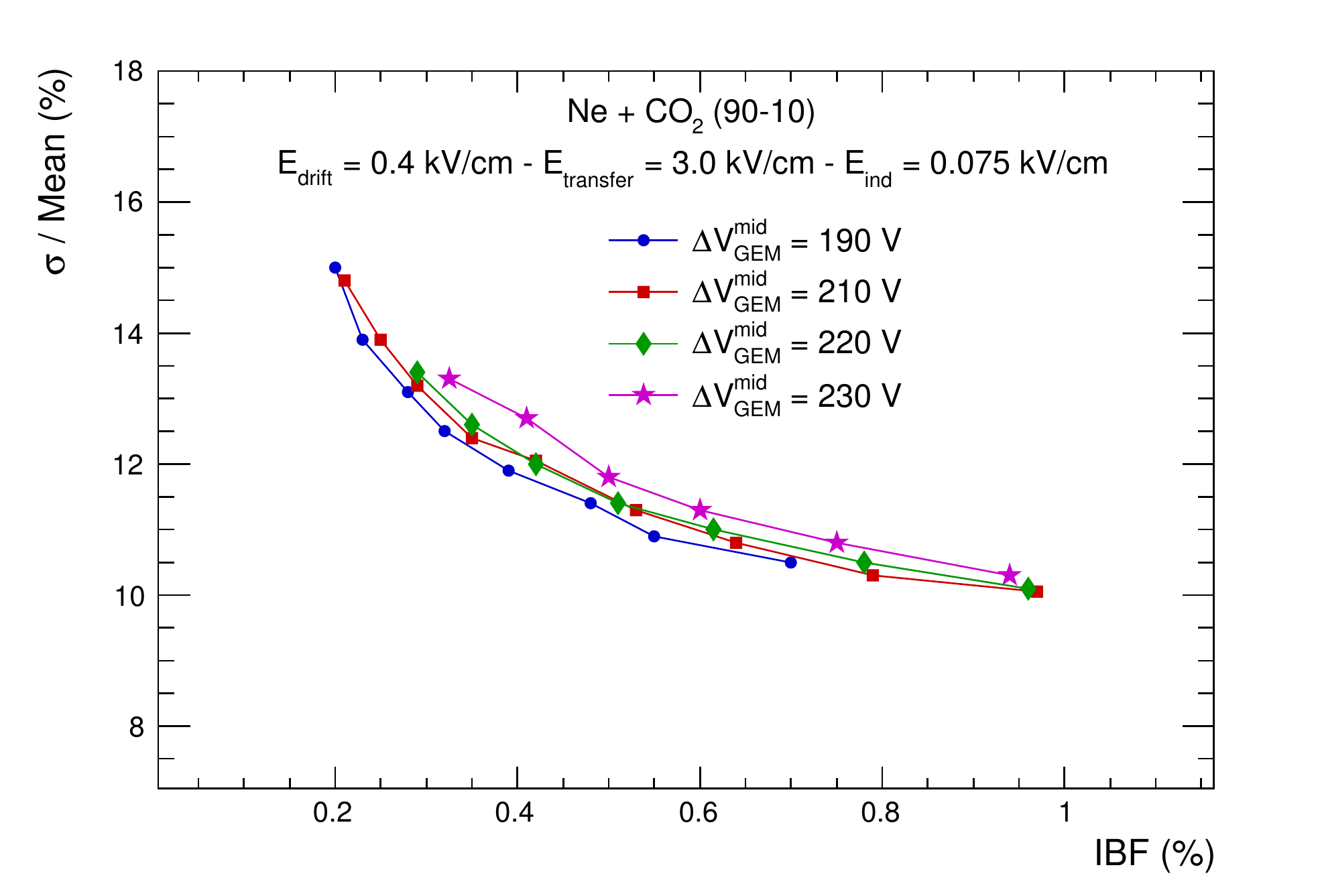}
\caption{Energy resolution vs. IBF, varying {\VMMG} and {\VGEMOne} for several fixed values of {\VGEMTwo}.} 
\label{fig:NeCO2-dV-GEM2}
\end{center}
\end{figure}

In order to optimize the performance of the system, we examined these scans in the 2D phase space of 
energy resolution and IBF. Figure~\ref{fig:NeCO2-dV-GEM2} shows the result for a Ne+{\COTwo} (90--10) gas 
mixture. Energy resolution vs. IBF curves are shown for various fixed {\VGEMTwo}; we scanned {\VMMG}
with {\VGEMOne} tuned as necessary to keep the overall gain fixed at about 2000. The result of this procedure 
defines a curve in this 2D space, for each fixed {\VGEMTwo}. While there is not a large difference in performance, 
there is a slight preference for lower {\VGEMTwo}. 

\begin{figure}[htb!]
\centering
\begin{subfigure}[b]{0.9\textwidth}
\includegraphics[width=\textwidth]{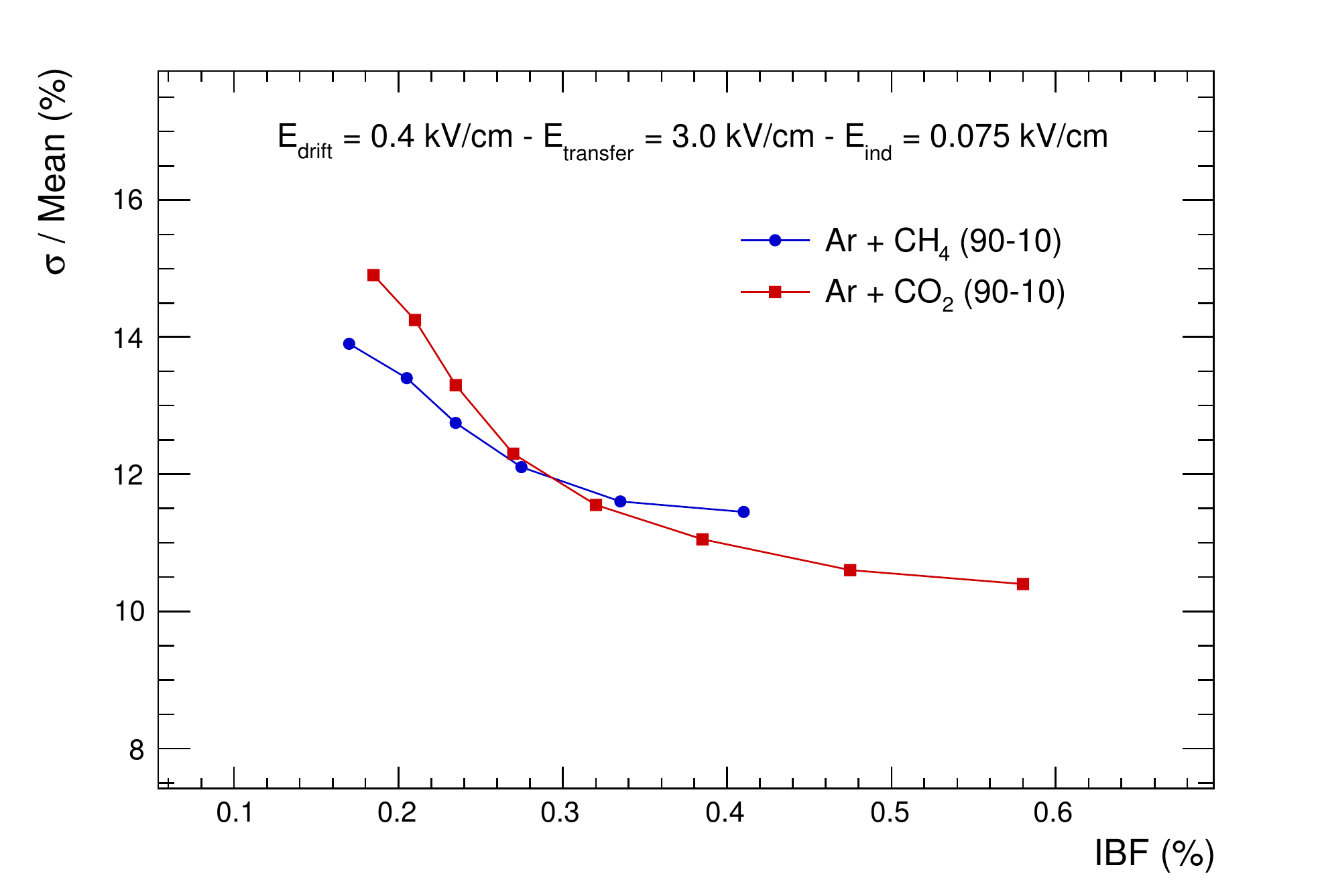}
\caption{Argon-based gas mixtures.}
\label{fig:GasMixturesAr_ResVsIbf}
\end{subfigure}

\begin{subfigure}[b]{0.9\textwidth}
\includegraphics[width=\textwidth]{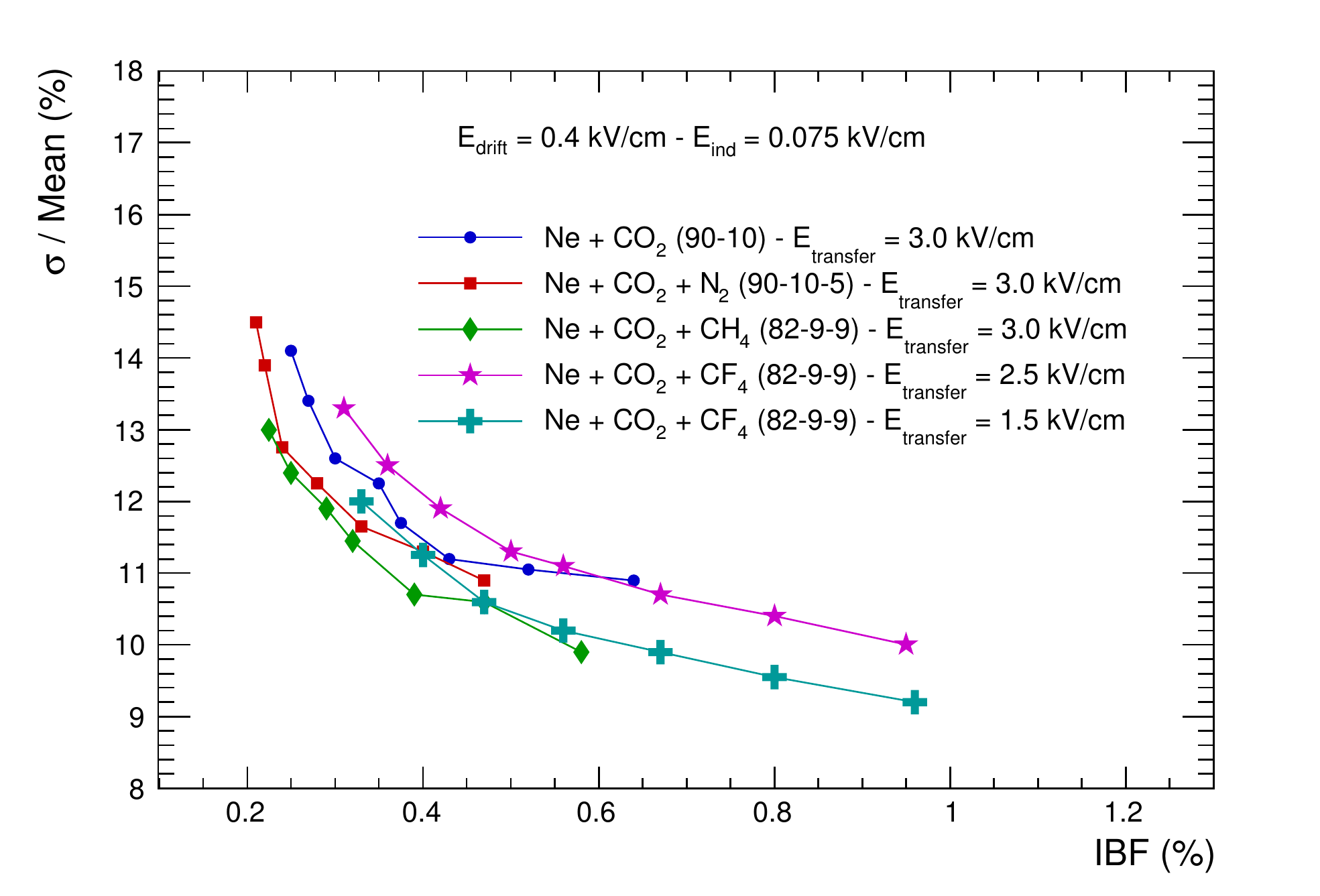}
\caption{Neon-based gas mixtures.}
\label{fig:GasMixturesNe_ResVsIbf}
\end{subfigure}

\caption{Energy resolution vs. IBF for a chamber with two stacked GEM foils and one MMG, filled with several gas mixtures based on neon or argon.}
\label{fig:GasMixtures_ResVsIbf}    
\end{figure}

For several sets of GEM voltages and electric fields we also measured the IBF for different {\VMMG} up to a 
chamber gain of 5500.  We found the product $\epsilon\equiv\text{IBF}\times\text{gain}$ is almost constant, as can be seen in 
Table~\ref{table:epsilon}. The increased ion production in the MMG as its gain increases is approximately 
compensated by the increased ion capture in the MMG due to the increase in the ratio of {\EMMG} to {\EInd}.

\begin{table}[hb!]
\centering
\begin{tabular}{llll}
    \toprule
    Gain & IBF (\%)  & $\epsilon$ & Gas Mixture      \\
    \midrule
    2000 & 0.3  & 6.0        & Ar+{\COTwo} (90--10) \\
    3000 & 0.21 & 6.3        & Ar+{\COTwo} (90--10) \\
    5500 & 0.11 & 6.5        & Ar+{\COTwo} (70--30) \\
    \bottomrule
\end{tabular}
\caption{Values of $\epsilon\equiv\text{IBF}\times\text{gain}$ extracted from the measurement of the gain and IBF with energy resolution fixed at 12\% along with the {\EDrift}, {\ETransfer}, and {\EInd} fixed. {\VGEMOne} and {\VGEMTwo} were approximately the same, although the precise setup and foils varied.}
\label{table:epsilon}
\end{table}

Additionally, we performed the same energy resolution vs. IBF measurements for a variety of argon and neon based gas mixtures. 
Figure~\ref{fig:GasMixturesAr_ResVsIbf} shows a comparison between {\CHFour} and {\COTwo} 
in argon. The {\COTwo} mixture exhibits better energy resolution at slightly higher IBF. Figure~\ref{fig:GasMixturesNe_ResVsIbf} 
shows a comparison between a number of different gases added to the baseline Ne+{\COTwo} mixture. Note that {\ETransfer} is 
lower for {\CFFour} to avoid the electron capture described above. These measurements seem to suggest a slight preference 
for Ne+{\COTwo}+{\CHFour}.

Note that when designing a TPC, the relevant parameter determining SCD is the ion density in the main TPC drift volume, 
which depends not just on IBF but also on other parameters such as the ion mobility and the level of primary ionization. The curves from
Fig.~\ref{fig:NeCO2-dV-GEM2} and Fig.~\ref{fig:GasMixtures_ResVsIbf} should be interpreted accordingly; despite similar IBF curves
in neon and argon, an argon-based gas mixture would result in higher space-charge buildup, due to its smaller ion mobility.

\subsection{MMG crosstalk effect and E-field uniformity}

In a high-rate environment, crosstalk between readout elements degrades the energy resolution of a detector. For a MMG detector, the mesh is quite close 
to the readout plane. 
This means that the capacitance between the mesh and the readout elements is larger than is typical for wire chambers or GEM 
chambers, which leads to increased capacitive coupling between readout elements and thus increased crosstalk. In a large chamber 
operating at a high rate, this crosstalk can degrade the energy resolution. For the chambers we tested, with a 126 micron gap and 
400 lines per inch mesh, we measured a mesh to readout capacitance of $17\;\text{pF/cm}^2$.  In our small chamber tests, using 
a standard charge sensitive preamplifier and a shaping amplifier with a $2\;\mu\text{s}$ shaping time, we measured an inverse polarity crosstalk 
amplitude of $0.35\% \text{ per cm}^2$ pad size, with the expectation that the crosstalk is proportional to the readout pad to 
mesh capacitance. 

Another feature of MMG elements resulting from the small gap between the mesh and readout plane is the influence of the 
readout pattern on the energy resolution. The width of the spaces between readout elements will be a significant fraction 
of the gap to the mesh and will therefore cause large local variation in the electric field and hence the gain. For example, we 
tested a MMG with chevron style pad readout (6 zigzags, $4\times8 \mbox{ mm}^{2}$ pads), which showed 30\% worse energy 
resolution compared to the MMG with rectangular pads of the same size.

\subsection{Discharge Rate}

To test the discharge behavior, we constructed a spark test chamber with spark protection on the MMG and readout plane. 
The chamber had a larger drift gap (43.6 mm), with a collimated $^{241}$Am source 11.1 mm above a small hole in the cathode. This source 
could be remotely moved relative to the cathode in order to vary the rate of $\alpha$ particles in the chamber.
An $^{55}$Fe source was also mounted in the chamber to monitor the chamber gain. Signals from 
the two GEM foils and the MMG mesh were coupled through capacitors, attenuators, and discriminators to scalers to count 
sparks.  The signal from the anode was also coupled to a scaler to count the total number of $\alpha$ particles.

We measured a discharge rate of less than $3\times10^{-7}$ per $\alpha$ in Ne+{\COTwo} (90--10). With Ne+{\COTwo}+{\CHFour} 
(82--9--9), the discharge rate decreased by an order of magnitude ($2\times10^{-8}$).

To further measure the discharge behavior, two $21\times26$ cm$^2$ detectors assembled at Yale, with MMG produced at CERN 
and GEM foils from the PHENIX Hadron Blind Detector~\cite{Anderson11}, filled with Ne+{\COTwo}+{\NTwo} (90--10--5), were tested by 
the ALICE TPC-Upgrade Collaboration in a SPS beam at CERN~\cite{Lippmann15}. The beam of $6\times10^{6}$ 150 GeV pions per 
4.5 s spill was incident on a 40 cm iron target just upstream of the chambers, creating a high-intensity mixed particle shower 
perpendicular to the pad plane. The equivalent minimum ionizing particle flux incident on the chamber was measured by calibrating 
the anode current of a chamber just upstream of our test chambers to the counted beam flux without the iron target.  A 
discharge rate of $3.5\times10^{-10}$ per chamber particle was measured. Approximately $5\times10^{11}$ chamber particles were 
accumulated. It was also observed that the sparking rate does not change much when the GEM voltages and transfer fields are 
switched off, indicating that the sparking is mainly due to the interaction of beam particles with the MMG. It should be noted that the 
spark does not damage the MMG, but rather poses the problem of a high voltage drop (with resultant dead time) and risk for readout electronics. 
Work is in progress to improve spark protection, such as providing a resistive layer on the pad plane to limit the discharge, and the HV drop and
its recovery time~\cite{Bay02,Alexopoulos11,Bortfeld13}.

\section{Conclusions}

In an effort to eliminate the standard gating grid in TPCs by minimizing the buildup of space charge in the drift volume, 
we investigated the use of 2-GEM + MMG chambers for the TPC gas amplification region. 
We selected this combination of MPGDs with the intention of minimizing {\EInd}/{\EMMG} independent of the TPC drift field, 
while keeping good energy resolution. To achieve good energy resolution, we employed a strong transfer field between the foils, and 
operated the top GEM with 3--5 effective gas amplification. To achieve a low induction field, the mid GEM was used to transfer 
electrons from the strong field to the weak field, with effective gas amplification smaller than one. With this configuration, the GEM foils 
provide the necessary field structure, additional IBF suppression, gain pre-amplification, and additional 
electron spread over the MMG surface. We focused on neon-based gas mixtures. In general, TPC optimization is a multi-parameter problem; 
if the correction of SCDs is the main factor for spatial resolution and momentum reconstruction performance,
neon-based gas mixtures (without isobutane) are suitable due to their large ion mobility, large TPC drift field, and small primary ionization. 

We achieved simultaneously an ion back-flow below 0.4\% (with 10--15\% uncertainty) and an energy resolution better than {\sigmaOverE}
$\;= 12\%$ (with 3--5\% uncertainty) for $^{55}$Fe X-rays at a gain of $\approx2000$ in a variety of gas mixtures. We reported the dependence 
of ion back-flow and energy resolution on the various field and amplification voltages. We also presented results on crosstalk and 
sparking from bench tests and with test beams. The hybrid micro-pattern gas amplification stage allows for a TPC design that can 
operate in a continuous mode, and serves as a viable option to limit space charge distortions in high-rate TPCs.

\section{Acknowledgments}

We acknowledge the ALICE TPC-Upgrade team for help in setting up and operating our chambers at the CERN PS and SPS beams, as 
well as the PHENIX Hadron Blind Detector collaboration for supplying GEM foils and readout electronics.

This work is supported by the US Department of Energy under Grant DE-SC004168, contract 200935 from Brookhaven National 
Laboratory, primary funding from US Department of Energy DE-AC02-98-CH10886, and contract 4000132727 from Oak Ridge National 
Laboratory, primary funding from US Department of Energy award DE-SC0014550.

\appendix{}
\counterwithin{table}{section}

\section{Data tables}

\begin{longtable}{rrrrrr}
\captionsetup{font=normalsize}
\caption{Data corresponding to Fig.~\ref{fig:NeCO2_EdriftScan}.}
\label{table:NeCO2_EdriftScan-table} \\
    \toprule
    \multicolumn{6}{c}{{\VMMG} = 400 V --- {\VGEMOne} = 242 V --- {\VGEMTwo} = 185 V}                   \\*
    \multicolumn{6}{c}{All electric fields are in units of kV/cm}                                       \\* \midrule
    {\EDrift} & {\ETransfer} & {\EInd} & $\text{I}_{\text{anode}}$ (nA) & {\sigmaOverE} (\%) & IBF (\%) \\* \midrule
    0.220     & 2.000        &   0.075 &  71.0                          & 10.90              & 0.200    \\*
    0.265     & 2.000        &   0.075 &  72.0                          & 10.95              & 0.250    \\*
    0.311     & 2.000        &   0.075 &  72.8                          & 10.95              & 0.300    \\*
    0.355     & 2.000        &   0.075 &  73.0                          & 10.90              & 0.340    \\*
    0.400     & 2.000        &   0.075 &  73.7                          & 10.90              & 0.400    \\*
    0.445     & 2.000        &   0.075 &  74.1                          & 10.90              & 0.450    \\*
    0.490     & 2.000        &   0.075 &  74.5                          & 10.90              & 0.490    \\* \bottomrule
\end{longtable}

\begin{longtable}{rrrrrr}
\captionsetup{font=normalsize}
\caption{Data corresponding to Fig.~\ref{fig:NeCO2_EtranScan}.}
\label{table:NeCO2_EtranScan-table} \\
    \toprule
    \multicolumn{6}{c}{{\VMMG} = 400 V --- {\VGEMOne} = 242 V --- {\VGEMTwo} = 185 V}                   \\*
    \multicolumn{6}{c}{All electric fields are in units of kV/cm}                                       \\* \midrule
    {\EDrift} & {\ETransfer} & {\EInd} & $\text{I}_{\text{anode}}$ (nA) & {\sigmaOverE} (\%) & IBF (\%) \\* \midrule
    0.400     & 0.330        & 0.075   &  52.5                          & 11.60              &  0.700   \\*
    0.400     & 0.670        & 0.075   &  73.0                          & 10.65              &  0.630   \\*
    0.400     & 1.000        & 0.075   &  75.5                          & 10.70              &  0.560   \\*
    0.400     & 1.330        & 0.075   &  75.1                          & 10.70              &  0.490   \\*
    0.400     & 1.670        & 0.075   &  75.8                          & 10.90              &  0.440   \\*
    0.400     & 2.000        & 0.075   &  74.8                          & 11.20              &  0.410   \\*
    0.400     & 2.170        & 0.075   &  73.8                          & 11.40              &  0.400   \\*
    0.400     & 2.330        & 0.075   &  73.0                          & 11.50              &  0.390   \\*
    0.400     & 2.500        & 0.075   &  73.7                          & 11.60              &  0.380   \\*
    0.400     & 2.670        & 0.075   &  71.0                          & 11.70              &  0.380   \\*
    0.400     & 2.830        & 0.075   &  69.5                          & 11.80              &  0.380   \\*
    0.400     & 3.000        & 0.075   &  68.9                          & 11.95              &  0.380   \\*
    0.400     & 3.170        & 0.075   &  69.0                          & 11.90              &  0.375   \\*
    0.400     & 3.330        & 0.075   &  69.5                          & 12.00              &  0.375   \\*
    0.400     & 3.500        & 0.075   &  71.0                          & 11.95              &  0.370   \\* \bottomrule
\end{longtable}

\begin{longtable}{rrrrrr}
\captionsetup{font=normalsize}
\caption{Data corresponding to Fig.~\ref{fig:NeCO2CF4_EtranScan}.}
\label{table:NeCO2CF4_EtranScan-table} \\
    \toprule
    \multicolumn{6}{c}{{\VMMG} = 430 V --- {\VGEMOne} = 271 V --- {\VGEMTwo} = 206 V} \\*
    \multicolumn{6}{c}{All electric fields are in units of kV/cm}                     \\* \midrule
    {\EDrift} & {\ETransfer} & {\EInd} & Gain & {\sigmaOverE} (\%) & IBF (\%)         \\* \midrule
    0.400     &  0.330       & 0.075   &  961 & 11.80              & 1.060            \\*
    0.400     &  0.830       & 0.075   & 1760 & 10.50              & 0.860            \\*
    0.400     &  1.500       & 0.075   & 2000 &  9.90              & 0.670            \\*
    0.400     &  1.830       & 0.075   & 1909 & 10.00              & 0.630            \\*
    0.400     &  2.170       & 0.075   & 1795 & 10.10              & 0.600            \\*
    0.400     &  2.500       & 0.075   & 1620 & 10.40              & 0.580            \\*
    0.400     &  2.830       & 0.075   & 1348 & 10.80              & 0.610            \\*
    0.400     &  3.170       & 0.075   & 1035 & 11.70              & 0.680            \\* \bottomrule
\end{longtable}

\begin{longtable}{rrrrrr}
\captionsetup{font=normalsize}
\caption{Data corresponding to Fig.~\ref{fig:NeCO2_EindScan}.}
\label{table:NeCO2_EindScan-table} \\
    \toprule
    \multicolumn{5}{c}{{\EDrift} = 0.4 kV/cm --- {\ETransfer} = 2.0 kV/cm --- {\VMMG} = 400 V}       \\* \midrule
    {\EInd} (kV/cm) & {\VGEMOne} (V) & {\VGEMTwo} (V) & {\sigmaOverE} (\%) & IBF (\%)                \\* \midrule
    0.025           & 242            & 215            & 10.90              & 0.630                   \\*
    0.050           & 242            & 195            & 10.70              & 0.470                   \\*
    0.075           & 242            & 185            & 10.70              & 0.400                   \\*
    0.125           & 242            & 171            & 11.20              & 0.350                   \\*
    0.187           & 232            & 168            & 11.85              & 0.320                   \\*
    0.250           & 220            & 168            & 12.30              & 0.320                   \\* \bottomrule
\end{longtable}

\begin{longtable}{rrrrrr}
\captionsetup{font=normalsize}
\caption{Data corresponding to Fig.~\ref{fig:NeCO2-one-scan}.}
\label{table:NeCO2-one-scan-table} \\
    \toprule
    \multicolumn{6}{c}{{\EDrift} = 0.4 kV/cm --- {\ETransfer} = 3 kV/cm --- {\EInd} = 0.075 kV/cm}   \\* \midrule
    {\VMMG} (V)       & {\VGEMOne} (V) & {\VGEMTwo} (V) & MMG GA & {\sigmaOverE} (\%) & IBF (\%)       \\* \midrule
    380               & 285            & 234            &  275   &  9.00              & 1.130          \\*
    390               & 280            & 227            &  343   &  9.00              & 0.910          \\*
    400               & 275            & 221            &  427   &  9.10              & 0.740          \\*
    410               & 268            & 214            &  534   &  9.50              & 0.610          \\*
    420               & 261            & 207            &  673   &  9.90              & 0.510          \\*
    430               & 254            & 204            &  846   & 10.10              & 0.420          \\*
    440               & 246            & 200            & 1066   & 10.70              & 0.340          \\*
    450               & 239            & 193            & 1347   & 11.50              & 0.300          \\*
    460               & 230            & 192            & 1709   & 12.80              & 0.250          \\* \bottomrule
\end{longtable}

\begin{longtable}{rrrrrr}
\captionsetup{font=normalsize}
\caption{Data corresponding to Fig.~\ref{fig:NeCO2-dV-GEM2}.}
\label{table:NeCO2-dV-GEM2-table} \\
    \toprule
    \multicolumn{6}{c}{{\VGEMTwo} = 190 V}                                                     \\* \midrule
    {\VMMG} (V)       & {\VGEMOne} (V) & {\VGEMTwo} (V) & Gain & {\sigmaOverE} (\%) & IBF (\%) \\* \midrule
    360               & 255            & 190            & 1874 & 10.50              & 0.700    \\*
    370               & 245            & 190            & 1905 & 10.90              & 0.550    \\*
    380               & 235            & 190            & 1926 & 11.40              & 0.480    \\*
    390               & 225            & 190            & 1958 & 11.90              & 0.390    \\*
    400               & 215            & 190            & 1984 & 12.50              & 0.320    \\*
    410               & 205            & 190            & 1905 & 13.10              & 0.280    \\*
    420               & 195            & 190            & 1941 & 13.90              & 0.230    \\*
    430               & 185            & 190            & 1953 & 15.00              & 0.200    \\ \midrule \nopagebreak
    \multicolumn{6}{c}{{\VGEMTwo} = 210 V}                                                     \\* \midrule
    {\VMMG} (V)       & {\VGEMOne} (V) & {\VGEMTwo} (V) & Gain & {\sigmaOverE} (\%) & IBF (\%) \\* \midrule
    340               & 255            & 210            & 1993 & 10.05              & 0.970    \\*
    350               & 245            & 210            & 1995 & 10.30              & 0.790    \\*
    360               & 235            & 210            & 2005 & 10.80              & 0.640    \\*
    370               & 225            & 210            & 1921 & 11.30              & 0.530    \\*
    380               & 215            & 210            & 1905 & 12.05              & 0.420    \\*
    390               & 205            & 210            & 1937 & 12.40              & 0.350    \\*
    400               & 195            & 210            & 1958 & 13.20              & 0.290    \\*
    410               & 185            & 210            & 1974 & 13.90              & 0.250    \\*
    420               & 175            & 210            & 1995 & 14.80              & 0.210    \\ \midrule \nopagebreak
    \multicolumn{6}{c}{{\VGEMTwo} = 220 V}                                                     \\* \midrule
    {\VMMG} (V)       & {\VGEMOne} (V) & {\VGEMTwo} (V) & Gain & {\sigmaOverE} (\%) & IBF (\%) \\* \midrule
    340               & 245            & 220            & 1869 & 10.10              & 0.960    \\*
    350               & 235            & 220            & 1877 & 10.50              & 0.780    \\*
    360               & 225            & 220            & 1888 & 11.00              & 0.615    \\*
    370               & 215            & 220            & 1899 & 11.40              & 0.510    \\*
    380               & 205            & 220            & 1940 & 12.00              & 0.420    \\*
    390               & 195            & 220            & 1971 & 12.60              & 0.350    \\*
    400               & 185            & 220            & 1985 & 13.40              & 0.290    \\ \midrule \nopagebreak
    \multicolumn{6}{c}{{\VGEMTwo} = 230 V}                                                     \\* \midrule
    {\VMMG} (V)       & {\VGEMOne} (V) & {\VGEMTwo} (V) & Gain & {\sigmaOverE} (\%) & IBF (\%) \\* \midrule
    340               & 235            & 230            & 1897 & 10.30              & 0.940    \\*
    350               & 225            & 230            & 1903 & 10.80              & 0.750    \\*
    360               & 215            & 230            & 1909 & 11.30              & 0.600    \\*
    370               & 205            & 230            & 1995 & 11.80              & 0.500    \\*
    380               & 195            & 230            & 1997 & 12.70              & 0.410    \\*
    390               & 185            & 230            & 2008 & 13.30              & 0.325    \\* \bottomrule
\end{longtable}

\begin{longtable}{rrrrrr}
\captionsetup{font=normalsize}
\caption{Data corresponding to Fig.~\ref{fig:GasMixturesAr_ResVsIbf}.}
\label{table:GasMixturesAr_ResVsIbf-table} \\
    \toprule
    \multicolumn{6}{c}{Ar+{\COTwo} (90--10)}                                                       \\*
    \multicolumn{6}{c}{{\EDrift} = 0.4 kV/cm --- {\ETransfer} = 3 kV/cm --- {\EInd} = 0.075 kV/cm} \\* \midrule
    {\VMMG} (V)       & {\VGEMOne} (V) & {\VGEMTwo} (V) & Gain & {\sigmaOverE} (\%) & IBF (\%)     \\* \midrule
    440               & 305            & 270            & 2248 & 10.40              & 0.580        \\*
    450               & 295            & 265            & 2074 & 10.60              & 0.475        \\*
    460               & 290            & 260            & 2146 & 11.05              & 0.385        \\*
    470               & 280            & 255            & 2046 & 11.55              & 0.320        \\*
    480               & 275            & 250            & 2092 & 12.30              & 0.270        \\*
    490               & 270            & 245            & 2146 & 13.30              & 0.235        \\*
    500               & 265            & 237            & 2111 & 14.25              & 0.210        \\*
    510               & 260            & 229            & 2091 & 14.90              & 0.185        \\ \toprule \nopagebreak
    \multicolumn{6}{c}{Ar+{\CHFour} (90--10)}                                                      \\*
    \multicolumn{6}{c}{{\EDrift} = 0.4 kV/cm --- {\ETransfer} = 3 kV/cm --- {\EInd} = 0.075 kV/cm} \\* \midrule
    {\VMMG} (V)       & {\VGEMOne} (V) & {\VGEMTwo} (V) & Gain & {\sigmaOverE} (\%) & IBF (\%)     \\* \midrule
    440               & 300            & 270            & 2175 & 11.45              & 0.410        \\*
    450               & 295            & 265            & 2229 & 11.60              & 0.335        \\*
    460               & 290            & 260            & 2320 & 12.10              & 0.275        \\*
    470               & 280            & 255            & 2099 & 12.75              & 0.235        \\*
    480               & 275            & 250            & 2168 & 13.40              & 0.205        \\*
    490               & 270            & 245            & 2242 & 13.90              & 0.170        \\* \bottomrule
\end{longtable}

\begin{longtable}{rrrrrr}
\captionsetup{font=normalsize}
\caption{Data corresponding to Fig.~\ref{fig:GasMixturesNe_ResVsIbf}.}
\label{table:GasMixturesNe_ResVsIbf-table} \\
    \toprule
    \multicolumn{6}{c}{Ne+{\COTwo} (90--10)}                                                         \\*
    \multicolumn{6}{c}{{\EDrift} = 0.4 kV/cm --- {\ETransfer} = 3 kV/cm --- {\EInd} = 0.075 kV/cm}   \\* \midrule
    {\VMMG} (V)       & {\VGEMOne} (V) & {\VGEMTwo} (V) & Gain & {\sigmaOverE} (\%) & IBF (\%)       \\* \midrule
    360               & 235            & 210            & 2055 & 10.90              & 0.640          \\*
    370               & 225            & 210            & 2140 & 11.05              & 0.520          \\*
    380               & 220            & 205            & 2151 & 11.20              & 0.430          \\*
    390               & 200            & 215            & 2142 & 11.70              & 0.375          \\*
    400               & 195            & 210            & 2133 & 12.25              & 0.350          \\*
    410               & 190            & 200            & 2134 & 12.60              & 0.300          \\*
    420               & 195            & 185            & 2160 & 13.40              & 0.270          \\*
    430               & 190            & 180            & 2163 & 14.10              & 0.250          \\ \toprule \nopagebreak
    \multicolumn{6}{c}{Ne+{\COTwo}+{\NTwo} (90--10--5)}                                              \\*
    \multicolumn{6}{c}{{\EDrift} = 0.4 kV/cm --- {\ETransfer} = 3 kV/cm --- {\EInd} = 0.075 kV/cm}   \\* \midrule
    {\VMMG} (V)       & {\VGEMOne} (V) & {\VGEMTwo} (V) & Gain & {\sigmaOverE} (\%) & IBF (\%)       \\* \midrule
    435               & 265            & 230            & 2057 & 10.90              & 0.470          \\*
    445               & 260            & 225            & 2090 & 11.30              & 0.400          \\*
    455               & 255            & 220            & 2126 & 11.65              & 0.330          \\*
    465               & 250            & 215            & 2172 & 12.25              & 0.280          \\*
    475               & 245            & 210            & 2223 & 12.75              & 0.240          \\*
    485               & 240            & 200            & 2049 & 13.90              & 0.220          \\*
    495               & 235            & 195            & 2102 & 14.50              & 0.210          \\ \toprule \nopagebreak
    \multicolumn{6}{c}{Ne+{\COTwo}+{\CHFour} (82--9--9)}                                           \\*
    \multicolumn{6}{c}{{\EDrift} = 0.4 kV/cm --- {\ETransfer} = 3 kV/cm --- {\EInd} = 0.075 kV/cm}   \\* \midrule
    {\VMMG} (V)       & {\VGEMOne} (V) & {\VGEMTwo} (V) & Gain & {\sigmaOverE} (\%) & IBF (\%)       \\* \midrule
    400               & 270            & 218            & 1985 & 9.90               & 0.580          \\*
    415               & 265            & 215            & 2122 & 10.60              & 0.470          \\*
    425               & 260            & 210            & 2160 & 10.70              & 0.390          \\*
    435               & 250            & 205            & 1969 & 11.45              & 0.320          \\*
    445               & 245            & 200            & 2004 & 11.90              & 0.290          \\*
    455               & 240            & 195            & 2044 & 12.40              & 0.250          \\*
    460               & 235            & 195            & 2055 & 13.00              & 0.225          \\ \toprule \nopagebreak
    \multicolumn{6}{c}{Ne+{\COTwo}+{\CFFour} (82--9--9)}                                           \\*
    \multicolumn{6}{c}{{\EDrift} = 0.4 kV/cm --- {\ETransfer} = 2.5 kV/cm --- {\EInd} = 0.075 kV/cm} \\* \midrule
    {\VMMG} (V)       & {\VGEMOne} (V) & {\VGEMTwo} (V) & Gain & {\sigmaOverE} (\%) & IBF (\%)       \\* \midrule
    405               & 325            & 250            & 1886 & 10.00              & 0.950          \\*
    415               & 320            & 245            & 1908 & 10.40              & 0.800          \\*
    425               & 315            & 240            & 1936 & 10.70              & 0.670          \\*
    435               & 310            & 235            & 1965 & 11.10              & 0.560          \\*
    445               & 305            & 230            & 2010 & 11.30              & 0.500          \\*
    455               & 300            & 225            & 1984 & 11.90              & 0.420          \\*
    465               & 295            & 220            & 2008 & 12.50              & 0.360          \\*
    475               & 290            & 215            & 2030 & 13.30              & 0.310          \\ \toprule \nopagebreak
    \multicolumn{6}{c}{Ne+{\COTwo}+{\CFFour} (82--9--9)}                                           \\*
    \multicolumn{6}{c}{{\EDrift} = 0.4 kV/cm --- {\ETransfer} = 1.5 kV/cm --- {\EInd} = 0.075 kV/cm} \\* \midrule
    {\VMMG} (V)       & {\VGEMOne} (V) & {\VGEMTwo} (V) & Gain & {\sigmaOverE} (\%) & IBF (\%)       \\* \midrule
    410               & 281            & 222            & 2361 & 9.20               & 0.960          \\*
    420               & 276            & 214            & 2324 & 9.55               & 0.800          \\*
    430               & 271            & 206            & 2281 & 9.90               & 0.670          \\*
    440               & 267            & 198            & 2275 & 10.20              & 0.560          \\*
    450               & 257            & 198            & 2385 & 10.60              & 0.470          \\*
    460               & 247            & 196            & 2403 & 11.25              & 0.400          \\*
    470               & 237            & 196            & 2507 & 12.00              & 0.330          \\* \bottomrule
\end{longtable}



\bibliography{yaleNIM}

\begin{thebibliography}{10}
\expandafter\ifx\csname url\endcsname\relax
  \def\url#1{\texttt{#1}}\fi
\expandafter\ifx\csname urlprefix\endcsname\relax\def\urlprefix{URL }\fi
\expandafter\ifx\csname href\endcsname\relax
  \def\href#1#2{#2} \def\path#1{#1}\fi

\bibitem{VanBuren06}
G.~V. Buren, et~al., Correcting for distortions due to ionization in the {STAR}
  {TPC}, NIM A566 (2006) 22.
\newblock \href {http://dx.doi.org/10.1016/j.nima.2006.05.131}
  {\path{doi:10.1016/j.nima.2006.05.131}}.

\bibitem{Blum08}
W.~Blum, W.~Riegler, L.~Rolandi, Particle Detection with Drift Chambers, 2nd
  Edition, Particle Acceleration and Detection, Springer-Verlag Berlin
  Heidelberg, 2008.
\newblock \href {http://dx.doi.org/10.1007/978-3-540-76684-1}
  {\path{doi:10.1007/978-3-540-76684-1}}.

\bibitem{Amendolia85}
S.~Amendolia, et~al., Ion trapping properties of a synchronously gated time
  projection chamber, NIM A239 (1985) 192.
\newblock \href {http://dx.doi.org/10.1016/0168-9002(85)90714-4}
  {\path{doi:10.1016/0168-9002(85)90714-4}}.

\bibitem{Wieman14}
H.~Wieman,
  \href{https://wiki.bnl.gov/eic/upload/Alice_upgrade_gating_grid_idea.pdf}{Gating
  grid concept for {ALICE} {TPC} upgrade}, private correspondence (04 2014).
\newline\urlprefix\url{https://wiki.bnl.gov/eic/upload/Alice_upgrade_gating_grid_idea.pdf}

\bibitem{Mulligan15}
J.~Mulligan, \href{http://arxiv.org/abs/1603.05648}{Simulations of a
  multi-layer extended gating grid} (03 2016).
\newline\urlprefix\url{http://arxiv.org/abs/1603.05648}

\bibitem{Sauli06}
F.~Sauli, L.~Ropelewski, P.~Everaerts, Ion feedback suppression in time
  projection chambers, NIM A560 (2006) 269.
\newblock \href {http://dx.doi.org/10.1016/j.nima.2005.12.239}
  {\path{doi:10.1016/j.nima.2005.12.239}}.

\bibitem{Colas04}
P.~Colas, I.~Giomataris, V.~Lepeltier, Ion backflow in the micromegas {TPC} for
  the future linear collider, NIM A535 (2004) 226.
\newblock \href {http://dx.doi.org/10.1016/j.nima.2004.07.274}
  {\path{doi:10.1016/j.nima.2004.07.274}}.

\bibitem{Bondar03}
A.~Bondar, A.~Buzulutskov, L.~Shekhtman, A.~Vasiljev, Study of ion feedback in
  multi-{GEM} structures, NIM A496 (2003) 325.
\newblock \href {http://dx.doi.org/10.1016/S0168-9002(02)01763-1}
  {\path{doi:10.1016/S0168-9002(02)01763-1}}.

\bibitem{ALICE14}
{ALICE Collaboration}, \href{https://cds.cern.ch/record/1622286}{Technical
  design report for the upgrade of the {ALICE}}, ALICE-TDR-016.
\newline\urlprefix\url{https://cds.cern.ch/record/1622286}

\bibitem{ALICE15}
{ALICE Collaboration}, \href{http://cds.cern.ch/record/1984329}{Addendum to the
  technical design report for the upgrade of the {ALICE} time projection
  chamber} CERN-LHCC-2015-002.
\newline\urlprefix\url{http://cds.cern.ch/record/1984329}

\bibitem{Sauli97}
F.~Sauli, {GEM}: A new concept for electron amplification in gas detectors, NIM
  A386 (1997) 531.
\newblock \href {http://dx.doi.org/10.1016/S0168-9002(96)01172-2}
  {\path{doi:10.1016/S0168-9002(96)01172-2}}.

\bibitem{Giomataris96}
Y.~Giomataris, P.~Rebourgeard, J.~Robert, G.~Charpak, {MICROMEGAS}: a
  high-granularity position-sensitive gaseous detector for high particle-flux
  environments, NIM A376 (1996) 29.
\newblock \href {http://dx.doi.org/10.1016/0168-9002(96)00175-1}
  {\path{doi:10.1016/0168-9002(96)00175-1}}.

\bibitem{Neyret12}
D.~Neyret, et~al., New pixelized micromegas detector with low discharge rate
  for the {COMPASS} experiment, Journal of Instrumentation 7~(03) (2012)
  C03006.
\newblock \href {http://dx.doi.org/10.1088/1748-0221/7/03/C03006}
  {\path{doi:10.1088/1748-0221/7/03/C03006}}.

\bibitem{Anderson11}
W.~Anderson, et~al., Design, construction, operation and performance of a
  hadron blind detector for the {PHENIX} experiment, NIM A646~(1) (2011) 35.
\newblock \href {http://dx.doi.org/10.1016/j.nima.2011.04.015}
  {\path{doi:10.1016/j.nima.2011.04.015}}.

\bibitem{Lippmann15}
{C. Lippmann, et al.}, A continuous read-out tpc for the alice upgrade,
  presentation at Pisa 2015 meeting, Proceedings to be published in NIM (2015).

\bibitem{Bay02}
A.~Bay, et~al., Study of sparking in micromegas chambers, NIM A488 (2002) 162.
\newblock \href {http://dx.doi.org/10.1016/S0168-9002(02)00510-7}
  {\path{doi:10.1016/S0168-9002(02)00510-7}}.

\bibitem{Alexopoulos11}
T.~Alexopoulos, et~al., A spark-resistant bulk-micromegas chamber for high-rate
  applications, NIM A640 (2011) 110.
\newblock \href {http://dx.doi.org/10.1016/j.nima.2011.03.025}
  {\path{doi:10.1016/j.nima.2011.03.025}}.

\bibitem{Bortfeld13}
J.~Bortfeld,
  \href{https://indico.cern.ch/event/245535/session/4/contribution/5/attachments/420744/584246/rd51MW_jbortfeldt.pdf}{Development
  of micromegas detectors with novel floating strip anode} (04 2013).
\newline\urlprefix\url{https://indico.cern.ch/event/245535/session/4/contribution/5/attachments/420744/584246/rd51MW_jbortfeldt.pdf}

\end{thebibliography}

\end{document}